# Mejora de la exploración y la explotación de las heurísticas constructivas para el MLSTP

Sergio Consoli[1], José A. Moreno-Pérez[2], Nenad Mladenović[1], Kenneth Darby-Dowman[1]

*Resumen*—En este trabajo se proponen dos mejoras para aumentar la explotación y la exploración del clásico algoritmo constructivo MVCA para el problema del árbol generador etiquetado mínimo (*Minimum Labelling Spanning Tree Problem*; MLSTP). Se describe la aplicación de contrastes de hipótesis no paramétricos para contrastar tales mejoras. En el MLSTP se parte de un grafo conexo con aristas de distinto tipo y se trata de encontrar el árbol generador con las aristas más parecidas posible. Cada tipo de arista viene identificado por un color o etiqueta y el árbol generador óptimo es aquel que usa el menor número de colores. Los tiempos y soluciones obtenidas son comparables a los mejores resultados aparecidos en la literatura para el MLSTP.
*Palabras clave*— **Árbol Generador, Heurística constructiva, Análisis estadístico no paramétrico.**

## I. Introducción

En una red comunicaciones mixta o un sistema de transporte multimodal los enlaces pueden ser de distinto tipo por lo que no pueden ser usados por todos los operadores. Para conectar una serie de nodos es importante determinar el menor número de operadores de permiten conectarlos. Si cada operador es identificado por un color o etiqueta en las aristas en las que opera, el problema consiste en encontrar el menor número de colores cuyas aristas permiten conectar dichos nodos. El problema estándar en este contexto consistirá en, dado un grafo cuyas aristas están etiquetadas, encontrar el menor número de estas etiquetas cuyas aristas constituyen un grafo parcial conexo, o equivalentemente un árbol generador. Aplicaciones reales de estos problemas en el campo de las comunicaciones han sido ya publicadas ([15], [16]) donde las etiquetas o colores corresponden a distinto tipo de enlace.

Este problema, como muchos otros problemas de árbol generador aparecidos en la literatura, tiene un gran conjunto de aplicaciones reales y se ha probado que es NP-completo. El problema **MLSTP** (*Minimum Labelling Spanning Tree Problem*) se formula en los términos siguientes. Sea $G = (V,E,L)$ un grafo no dirigido conexo etiquetado, donde $V$ es un conjunto de $n$ vértices y $E$ es el conjunto de $m$ aristas, cada una de las cuales tiene una etiqueta dentro de un conjunto de $l$ colores o etiquetas $L$. El etiquetado de las aristas del grafo puede venir representado por una aplicación $f_L: E \rightarrow L$. Sea $L_T$ el conjunto de diferentes etiquetas del árbol $T$. El propósito del problema es encontrar el árbol $T$ que minimiza $|L_T|$ En el presente trabajo se utilizan técnicas estadísticas no paramétricas para analizar diversas estrategias constructivas con el objeto de incorporarlas a un procedimiento del tipo GRASP con arranque múltiple para la solución del problema.

El resto del trabajo se organiza de la siguiente manera. En la siguiente sección describimos los antecedentes del problema. En la sección 3 discutimos las mejoras en la estrategia constructiva para resolverlo. La siguiente sección describe la experiencia computacional realizada y el trabajo finaliza con unas conclusiones.

## II. Antecedentes

El problema fue introducido con una motivación en el campo de las telecomunicaciones por Chang y Leu [3] quienes prueban que es NP-duro y proponen un algoritmo constructivo que sigue la estrategia *greedy* del máximo cubrimiento de vértices. A pesar de que esta heurística no garantiza la optimalidad ni la factibilidad, las sucesivas correcciones o mejoras han seguido manteniendo su nombre (*maximum vertex covering algorithm*, MVCA). El algoritmo parte del grafo parcial sin aristas $H = (V,\emptyset)$ y sucesivamente incluye todas las aristas de la etiqueta que cubre más vértices aún no cubiertos. La solución heurística es cualquier árbol generador de este grafo resultante. El pseudocódigo viene en la figura siguiente:

---

**Algoritmo MVCA**
```
1. Sea C ← ∅ el conjunto de colores usados
   inicialmente vacío.
2. Sea H=(V,E(C)) el grafo parcial formado
   por todas las aristas cuyo color está en
   C; E(C) = {e∈E: L(e)∈C}.
3. Mientras existan nodos no cubiertos:
   3.1. Encontrar el color c∈(L-C) cuyas
        aristas cubren el mayor número de
        vértices no cubiertos por E(C).
   3.2. Añadir c al conjunto C de colores
        usados: C ← C ∪ {c}.
   3.3. Actualizar H añadiendo las aristas
        de color c: E_H ← E_H ∪ E({c}).
4. Sea T un árbol generador de H.
```
---

Fig. 1.    Algoritmo MVCA

En [3] se observó que este algoritmo puede acabar en un subgrafo parcial no conexo con lo que el paso 4 no puede ejecutarse. Esto ocurre, por ejemplo, en el grafo de la figura 2. En la parte izquierda de la figura se muestra el grafo inicial. La solución óptima está en la parte derecha de la figura y consta de dos colores o etiquetas: 2 y 3.

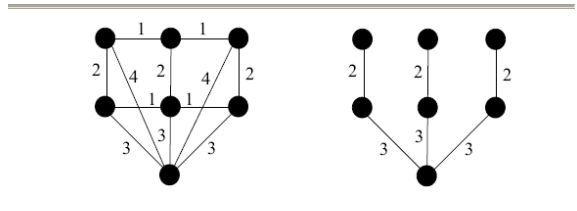

Fig. 2. Instancia y solución óptima del MLSTP

La ejecución del algoritmo MVCA en esta instancia se muestra en la figura 3. El procedimiento empieza eligiendo la etiqueta 1 que maximiza (junto con 2) el número de vértices no cubiertos (6). La siguiente iteración añade el color 3 que es el único que cubre un nuevo vértice y el algoritmo se detiene con el grafo parcial de la derecha de la figura 3 sin proporcionar ninguna solución factible.

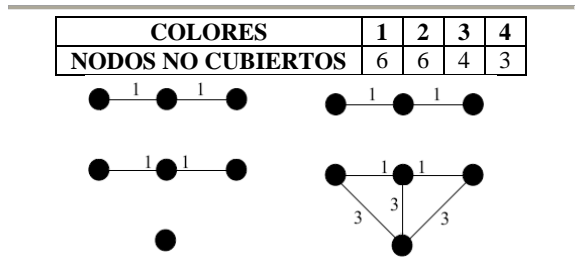

| COLORES | 1 | 2 | 3 | 4 |
|---|---|---|---|---|
| NODOS NO CUBIERTOS | 6 | 6 | 4 | 3 |

Fig. 3. Ejecución del MVCA original.

Este algoritmo fue corregido por Krumke y Wirth (1998) quienes proponen añadir en cada iteración el color que más reduce el número de componentes conexas del grafo parcial, en lugar de el que más nodos cubre. El proceso se detiene cuando el grafo parcial tenga una única componente conexa y de él se obtiene un árbol generador arbitrario. El nuevo algoritmo *greedy*, denominado **MVCA revisado**, está descrito en la figura 4.

El algoritmo MVCA revisado finaliza en una solución factible pero no necesariamente la óptima. Al aplicarlo al ejemplo mostrado en la figura 2, en las dos primeras iteraciones se seleccionan los mismos colores 1 y 3 que en el algoritmo sin revisar. Sin embargo, dado que el número de componentes conexas del grafo parcial resultante es 2, en una nueva iteración añadiría la etiqueta 2. El grafo resultante tiene tres etiquetas pero no corresponde a la solución óptima. En la ejecución del algoritmo, si varias etiquetas verifican el criterio de selección, se elige la primera que se encuentra.

Las otras elecciones posibles también dan lugar a una solución con tres colores.

**Algoritmo MVCA revisado**
```
1. Sea C ← ∅ el conjunto de colores usados
   inicialmente vacío.
2. Sea H=(V,E(C)) el grafo parcial formado
   por todas las aristas cuya color está en
   C; E(C) = {e∈E: L(e)∈C}.
3. Mientras el grafo H sea desconexo:
   3.1. Sea Comp(C) el número de compo-
        nentes conexas de H=(V,E(C)).
   3.2. Encontrar el color c∈(L-C) que
        minimiza Comp(C ∪ {c}).
   3.3. Añadir c al conjunto C de colores
        usados: C ← C ∪ {c}.
   3.4. Actualizar H añadiendo las aristas
        de color c: E_H ← E_H ∪ E({c}).
4. Sea T un árbol generador de of H.
```

Fig. 4. Algoritmo MVCA revisado

En la literatura se ofrecen varios resultados sobre acotaciones de la calidad de las soluciones obtenidas y la complejidad del problema. Krumke y Wirth [11] probaron que el algoritmo MVCA revisado aporta una solución no mayor que $(1 + 2 \log n)$ veces la óptima. Posteriormente en [17] se mejora esta cota mostrando que el algoritmo es una $(1+\log⊙(n−1))$-aproximación.

Por otro lado, en [1] se propuso aplicar búsquedas locales basadas en el $k$-intercambio a una versión restringida del MLSTP y se probaron varios resultados sobre complejidad demostrando que si cada etiqueta aparece a lo sumo dos veces el problema es resoluble en tiempo polinomial.

Las cotas de [17] y [11] no están ajustadas; no pueden alcanzarse nunca. En [19] se obtuvo un resultado ajustado. Primero probaron que si la frecuencia de las etiquetas está acotada por b, la cota del algoritmo MVCA del peor caso es el $b$-ésimo número harmónico $H_b = 1 + 1/2 + \ldots + 1/b$. Posteriormente construyeron una familia de grafos tales que la solución aportada por el algoritmo MVCA es exactamente $H_b$ veces la solución óptima. Se trata de una mejora de la cota propuesta en [18] ya que $H_b < (1 + \log⊙(n − 1))$ y $b \leq (n − 1)$.

La literatura especializada también ofrece otras propuestas heurísticas. En [19] se presenta un algoritmo genético que mejoraba la solución del MVCA en la mayoría de los casos. Posteriormente, en [2] aplican el método Piloto propuesto en [6]. El método Piloto combina una estrategia voraz o *greedy* en la selección del color con una capacidad limitada de análisis hacia adelante. Este procedimiento da lugar a muy buenas soluciones pero a costa de un tiempo de computación alto, sobre todo si hay muchas aristas. En [20] se proponen varias implementaciones del método de [2] con versiones simplificadas y más rápidas.

El algoritmo basado en el método Piloto propuesto en [2] toma como primera etiqueta del proceso constructivo cada uno de las *l* etiquetas posibles y luego prosigue con la estrategia *greedy*; selecciona el color que más reduce el número de componentes conexas. Una primera versión propuesta en [20] ordena las etiquetas en función de su frecuencia y prueba como primera etiqueta sólo el 10% de mayor frecuencia. Una vez elegida la primera etiqueta el algoritmo continúa con la estrategia *greedy*. El algoritmo selecciona la mejor solución de las $l/10$ así encontradas. El tiempo de ejecución se reduce en 90% con respecto al piloto y se pretende que la calidad de la solución aportada pueda no diferir significativamente en calidad. Una segunda versión aumenta hasta el 30% el porcentaje de colores con la mayor frecuencia probados como primera etiqueta. Además, en [20] se presenta una versión del algoritmo genético que compite las versiones modificadas del método Piloto aportando un compromiso entre la calidad del método más exhaustivo y el tiempo de cómputo del más rápido.

La solución exacta se puede obtener por un clásico método A* o *backtracking* que recorre los subconjuntos de *L*. El método de búsqueda A* ejecuta una estrategia inteligente de ramificación y poda en el espacio de las soluciones parciales basado en el procedimiento recursivo *Ensayar* que incorpora a la solución parcial un nuevo color. El programa principal llama al procedimiento *Ensayar* con la solución parcial vacía.

---

**Algoritmo A\***:
```
1. Sea C ← ∅ el conjunto de colores
   usados, inicialmente vacío.
2. Sea C* ← L el conjunto óptimo de
   colores, inicialmente completo.
3. Ensayar(C).
4. Sea T un árbol generador de
      H* = (V, E(C*)).
   Procedimiento Ensayar(C):
   Si |C| < |C*| entonces
      Si Comp(C) = 1 entonces C* ← C.
      En otro caso, si |C| < |C*|-1
         Entonces, para cada c ∈ (L-C):
               Ensayar(C ∪ {c}).
```

Fig. 5.     Algoritmo A* para el MLST

---

Usando como solución inicial una buena solución heurística, como la proporcionada por una versión eficiente y efectiva del MVCA revisado, se reduce considerablemente el tiempo de cómputo. Otra mejora consiste en rechazar las soluciones parciales que no pueden ser completadas con éxito para obtener un grafo conexo. Si una solución parcial *C'* verifica $|C'|=|C^*|-2$ no es necesario comprobar los colores con una frecuencia (número de aristas que contiene) menor que $Comp(C')-1$.

El cálculo del número de componentes conexas se realiza con el método de etiquetado que numera las componentes conexas [13]. Este procedimiento permite obtener eficientemente el número de componentes conexas $Comp(C \cup \{c\})$ partiendo de la numeración de $Comp(C)$.

En [3] se propusieron y analizaron diversas versiones del algoritmo constructivo MVCA. En ellas se realizaban distintas combinaciones del uso de las frecuencias de cada una de las etiquetas, el número de componentes conexas y las estrategias pilotos. La conclusión más relevante fue que el uso de la frecuencia de los colores no aportaba ninguna ventaja significativa.

Al aplicar en cada paso del proceso constructivo, la regla de Krumke y Wirth de elegir el color que minimizaba el número de componentes conexas se aproxima rápidamente a una solución de alta Calidad. Dado que frecuentemente existen muchos colores que alcanzan ese mínimo, los resultados del algoritmo dependen en gran medida de la regla utilizada para elegir entre ellos. Si se toma el primer color que se encuentra, los resultados vienen condicionados por la ordenación previa de los colores. Por tanto, distintas ejecuciones del algoritmo podrían dar lugar a distintas soluciones, que incluso podrían tener distinto número de etiquetas.

La estrategia Piloto se aplicó explorando, en cada iteración, todos los colores que minimizan el número de componentes conexas en cada iteración. De esta forma se recorren todas las soluciones que es posible alcanzar con la regla de Krumke y Wirth, pero incluso así era imposible alcanzar la solución óptima en algunas instancias. Sin embargo, aplicando la estrategia Piloto que explora todos los colores en primer paso se consiguió alcanzar el óptimo en muchas de esas instancias.

Ambas propuestas de aplicación de la estrategia *Piloto* se implementaron modificando el algoritmo A* de la figura 5. El procedimiento recursivo *Ensayar*, en lugar de hacer llamadas recursivas con $C \cup \{c\}$ para cada color $c \in (L-C)$, sólo las realiza con los colores que minimizan $Comp(C \cup \{c\})$. Para explorar en la primera etapa la inclusión de todos los colores de *L*, se modifica el paso 3 del algoritmo A* de la figura 5. Por tanto el procedimiento que incorpora ambas aplicaciones de la estrategia Piloto es el descrito en la figura 6.

---

**Algoritmo Piloto**:
```
1. Sea C ← ∅ el conjunto de colores
   usados, inicialmente vacío.
2. Sea C* ← L el conjunto óptimo de
   colores, inicialmente completo.
3. Para cada c ∈ L: Ensayar({c}).
4. Sea T un árbol generador de
      H* = (V, E(C*)).
   Procedimiento Ensayar(C):
   Si |C| < |C*| entonces
      Si Comp(C) = 1 entonces C* ← C.
      En otro caso, si |C| < |C*|-1
         Entonces, para cada c ∈ (L-C)
         que minimice Comp(C ∪ {c}):
               Ensayar(C ∪ {c}).
```

Fig. 6.     Algoritmo A* para el MLST

III. ESTUDIO EXPERIMENTAL

El presente estudio experimental se diseñó con el objeto de proponer y analizar dos mejoras del algoritmo MVCA revisado tendentes a mejorar su capacidad de explotación y su capacidad de exploración, en combinación con un arranque múltiple. Se consideró que la versión más aséptica del MVCA revisado era la ejecución aleatorizada del criterio de Krumke y Wirth por lo que, en cada iteración, si existen varias etiquetas que alcanzan el mínimo de componentes conexas, se elige una de ellas al azar. Así, al ejecutar repetidamente el algoritmo en un contexto de arranque múltiple aumenta la probabilidad de alcanzar el óptimo.

En el ejemplo de las figuras 2 y 3 se observa que el grafo resultante de la aplicación del algoritmo MVCA revisado tiene tres etiquetas, mientras que la solución óptima tiene sólo dos. Ejecuciones distintas del algoritmo MVCA podrían dan lugar a otras soluciones con tres colores; que tampoco son óptimas. Si consideramos la solución aportada por la primera de las ejecuciones determinada por el conjunto $C = \{1, 2, 3\}$ se observa que es posible prescindir del color 1 y el grafo seguiría siendo conexo. Cuando finaliza el algoritmo MVCA revisado conviene comprobar si se puede eliminar algún color sin desconectar el grafo. Por tanto, la capacidad de explotación del algoritmo aumenta con el proceso de post-optimización consistente en eliminar iterativamente un color que mantenga el grafo conexo, mientras exista.

Por otro lado, la estrategia piloto que refuerza la capacidad de exploración del procedimiento consistió en explorar cada uno de los colores como primer color elegido. En un contexto de arranque múltiple, un objetivo similar se conseguiría eligiendo totalmente al azar el primer color en el proceso constructivo, en lugar de minimizando el número de componentes conexas.

*A. Algoritmos utilizados*

Para la experiencia computacional, además del *algoritmo A\** y las diferentes versiones de la estrategia piloto descritas, se implementó la versión aleatorizada del algoritmo MVCA de Krumke y Wirth que denotamos por *algoritmo A*. Denotamos por *algoritmo A1* a este algoritmo con la modificación consistente en incorporar el proceso de post-optimización anteriormente descrito y por *algoritmo A2* al consistente en incorporar la elección totalmente al azar del primer color. Finalmente, denotamos por *algoritmo A12* al resultante de incorporar ambas modificaciones del algoritmo. Más concretamente, el algoritmo A12 es el descrito en la figura 7. El algoritmo A1 se obtiene al suprimir el paso 5 y el algoritmo A2 al suprimir el paso 2. Finalmente, si se suprimen los pasos 2 y 5 se obtiene el algoritmo A.

**Algoritmo A12**
1. Sea $C \leftarrow \emptyset$ el conjunto de colores usados inicialmente vacío.
2. Elegir un color $c$ al azar y añadirlo al conjunto inicial $C$; $C = \{c\}$.
3. Sea $H=(V,E(C))$ el grafo parcial formado por todas las aristas cuya color está en $C$; $E(C) = \{e \in E: L(e) \in C\}$.
4. Mientras el grafo $H$ sea desconexo:
   4.1. Sea $Comp(C)$ el número de componentes conexas de $H=(V,E(C))$.
   4.2. Elegir al azar un color $c \in (L-C)$ que minimice $Comp(C \cup \{c\})$.
   4.3. Añadir $c$ al conjunto $C$ de colores usados: $C \leftarrow C \cup \{c\}$.
   4.4. Actualizar $H$ añadiendo las aristas de color $c$: $E_H \leftarrow E_H \cup E(\{c\})$.
5. Post-optimización
   5.1. Buscar un color $c \in C$ tal que: $Comp(C / \{c\}) = 1$
   5.2. Si no existe el color $c$ ir a 4.4
   5.3. Hacer $C \leftarrow C / \{c\}$ y volver a 4.1.
   5.4. Sea $H=(V,E(C))$
6. Sea $T$ un árbol generador de of $H$.

Fig. 7.    Algoritmo A12.

*B. Instancias*

Para el estudio experimental se han usado varios conjuntos de instancias generados aleatoriamente atendiendo a diversas características que influyen en la dificultad de su solución. Siguiendo las pruebas realizadas en trabajos previos, las instancias se generan en base a tres parámetros: el *número n* de vértices, el número *l* de etiquetas o colores y la densidad *d* del grafo medida como el cociente entre el número *m* de aristas y el número de aristas del grafo completo de *n* vértices: $n(n-1)/2$. El número de vértices y colores se eligen iguales y variando entre 20 y 50 de 10 en 10. Se consideran tres escenarios con grafos de alta, media o baja densidad correspondiente a los valores $d = 0,8$, $d = 0,5$ y $d = 0,2$. El número de aristas de cada instancias se fija a partir de la densidad y del número de vértices mediante $m = \lfloor d \cdot n(n-1)/2 \rfloor$.

Agradecemos a Cerulli R., et al. (2005) la cesión de los datos usados en nuestros experimentos; de esta forma los valores objetivos alcanzados son comparables directamente con sus resultados.

La tabla 1 muestra los promedios de los valores objetivos alcanzados al aplicar 100 veces los distintos algoritmos en las diferentes instancias. Las tres primeras columnas aportan los datos identificativos de la instancia y la cuarta los valores óptimos que fueron obtenidos con el algoritmo A\*. En las siguientes columnas se encuentran los valores promedios alcanzados por cada uno de los algoritmos, A, A1, A2 y A12. Los valores medios del número de colores o etiquetas en las 100 ejecuciones en todas las instancias son

| A* | A1 | A2 | A3 | A4 |
|---|---|---|---|---|
| 4.64 | 4.98 | 4.93 | 5.34 | 5.14 |

TABLA I
VALORES OBJETIVO PROMEDIO ALCANZADOS POR CADA ALGORITMO

| n | d | i | A* | A | A1 | A2 | A12 | n | d | i | A* | A | A1 | A2 | A12 |
|---|---|---|---|---|---|---|---|---|---|---|---|---|---|---|---|
| 20 | 0.8 | 1 | 3 | 3.00 | 3.00 | 3.08 | 3.03 | 40 | 0.8 | 1 | 3 | 3.00 | 3.00 | 3.21 | 3.15 |
| | | 2 | 2 | 3.00 | 2.98 | 2.90 | 2.87 | | | 2 | 3 | 3.00 | 3.00 | 3.26 | 3.22 |
| | | 3 | 2 | 2.00 | 2.00 | 2.75 | 2.49 | | | 3 | 3 | 3.00 | 3.00 | 3.29 | 3.24 |
| | | 4 | 2 | 2.75 | 2.73 | 2.86 | 2.76 | | | 4 | 3 | 3.00 | 3.00 | 3.05 | 3.04 |
| | | 5 | 2 | 2.00 | 2.00 | 2.85 | 2.66 | | | 5 | 3 | 3.00 | 3.00 | 3.24 | 3.16 |
| | | 6 | 2 | 2.00 | 2.00 | 2.62 | 2.37 | | | 6 | 2 | 2.00 | 2.00 | 2.92 | 2.61 |
| | | 7 | 3 | 3.00 | 3.00 | 3.00 | 3.00 | | | 7 | 3 | 3.00 | 3.00 | 3.09 | 3.05 |
| | | 8 | 2 | 2.00 | 2.00 | 2.85 | 2.77 | | | 8 | 3 | 3.00 | 3.00 | 3.15 | 3.03 |
| | | 9 | 3 | 3.00 | 3.00 | 3.07 | 3.04 | | | 9 | 3 | 3.00 | 3.00 | 3.40 | 3.31 |
| | | 10 | 3 | 3.00 | 3.00 | 3.06 | 3.04 | | | 10 | 3 | 3.00 | 3.00 | 3.23 | 3.11 |
| | 0.5 | 1 | 3 | 4.00 | 3.95 | 4.01 | 3.87 | | 0.5 | 1 | 3 | 4.00 | 3.99 | 4.30 | 4.17 |
| | | 2 | 3 | 3.00 | 3.00 | 3.60 | 3.22 | | | 2 | 4 | 4.36 | 4.32 | 4.40 | 4.37 |
| | | 3 | 3 | 3.00 | 3.00 | 3.69 | 3.37 | | | 3 | 4 | 4.26 | 4.25 | 4.56 | 4.49 |
| | | 4 | 3 | 3.44 | 3.34 | 3.71 | 3.62 | | | 4 | 4 | 4.00 | 4.00 | 4.27 | 4.10 |
| | | 5 | 3 | 3.52 | 3.43 | 3.86 | 3.71 | | | 5 | 4 | 4.00 | 4.00 | 4.68 | 4.47 |
| | | 6 | 3 | 4.00 | 3.80 | 3.81 | 3.73 | | | 6 | 4 | 4.00 | 4.00 | 4.74 | 4.58 |
| | | 7 | 4 | 4.00 | 4.00 | 4.00 | 4.00 | | | 7 | 3 | 3.00 | 3.00 | 3.96 | 3.76 |
| | | 8 | 3 | 4.00 | 4.00 | 4.11 | 4.00 | | | 8 | 3 | 4.00 | 4.00 | 4.01 | 3.88 |
| | | 9 | 3 | 3.00 | 3.00 | 3.79 | 3.55 | | | 9 | 4 | 4.00 | 4.00 | 4.45 | 4.36 |
| | | 10 | 3 | 4.00 | 4.00 | 4.01 | 3.96 | | | 10 | 4 | 4.35 | 4.35 | 4.99 | 4.81 |
| | 0.2 | 1 | 5 | 5.00 | 5.00 | 5.67 | 5.31 | | 0.2 | 1 | 7 | 7.40 | 7.19 | 8.30 | 7.86 |
| | | 2 | 6 | 6.64 | 6.57 | 7.08 | 6.63 | | | 2 | 7 | 7.39 | 7.32 | 8.15 | 7.62 |
| | | 3 | 7 | 7.76 | 7.65 | 8.13 | 7.76 | | | 3 | 8 | 9.71 | 9.41 | 9.76 | 9.47 |
| | | 4 | 7 | 7.54 | 7.00 | 7.86 | 7.01 | | | 4 | 8 | 10.02 | 9.81 | 9.95 | 9.62 |
| | | 5 | 5 | 5.26 | 5.24 | 5.86 | 5.50 | | | 5 | 7 | 8.61 | 8.41 | 9.09 | 8.67 |
| | | 6 | 7 | 7.20 | 7.11 | 7.53 | 7.17 | | | 6 | 8 | 9.19 | 8.87 | 9.54 | 9.00 |
| | | 7 | 7 | 7.07 | 7.06 | 7.48 | 7.17 | | | 7 | 7 | 7.78 | 7.75 | 8.24 | 7.91 |
| | | 8 | 8 | 8.01 | 8.01 | 8.40 | 8.12 | | | 8 | 7 | 7.38 | 7.29 | 7.82 | 7.56 |
| | | 9 | 8 | 8.21 | 8.09 | 8.46 | 8.09 | | | 9 | 7 | 7.24 | 7.21 | 7.75 | 7.46 |
| | | 10 | 7 | 7.36 | 7.31 | 7.90 | 7.49 | | | 10 | 8 | 8.87 | 8.84 | 9.41 | 9.03 |
| 30 | 0.8 | 1 | 3 | 3.00 | 3.00 | 3.11 | 3.04 | 50 | 0.8 | 1 | 3 | 3.00 | 3.00 | 3.25 | 3.20 |
| | | 2 | 2 | 2.00 | 2.00 | 2.92 | 2.66 | | | 2 | 3 | 3.00 | 3.00 | 3.49 | 3.46 |
| | | 3 | 3 | 3.00 | 3.00 | 3.04 | 3.01 | | | 3 | 3 | 3.00 | 3.00 | 3.39 | 3.31 |
| | | 4 | 3 | 3.00 | 3.00 | 3.00 | 3.00 | | | 4 | 3 | 3.00 | 3.00 | 3.19 | 3.16 |
| | | 5 | 3 | 3.00 | 3.00 | 3.08 | 3.04 | | | 5 | 3 | 3.00 | 3.00 | 3.22 | 3.19 |
| | | 6 | 3 | 3.00 | 3.00 | 3.05 | 3.00 | | | 6 | 3 | 3.00 | 3.00 | 3.26 | 3.22 |
| | | 7 | 2 | 2.00 | 2.00 | 2.83 | 2.74 | | | 7 | 3 | 3.00 | 3.00 | 3.33 | 3.23 |
| | | 8 | 3 | 3.00 | 3.00 | 3.15 | 3.07 | | | 8 | 3 | 3.00 | 3.00 | 3.33 | 3.30 |
| | | 9 | 3 | 3.00 | 3.00 | 3.00 | 3.00 | | | 9 | 3 | 3.00 | 3.00 | 3.28 | 3.25 |
| | | 10 | 3 | 3.00 | 3.00 | 3.04 | 3.00 | | | 10 | 3 | 3.00 | 3.00 | 3.24 | 3.19 |
| | 0.5 | 1 | 4 | 4.00 | 4.00 | 4.15 | 4.07 | | 0.5 | 1 | 4 | 4.00 | 4.00 | 4.32 | 4.27 |
| | | 2 | 4 | 4.14 | 4.13 | 4.31 | 4.23 | | | 2 | 4 | 4.00 | 4.00 | 4.72 | 4.53 |
| | | 3 | 3 | 3.41 | 3.39 | 3.96 | 3.66 | | | 3 | 4 | 5.00 | 5.00 | 5.28 | 5.18 |
| | | 4 | 3 | 3.00 | 3.00 | 3.71 | 3.50 | | | 4 | 4 | 4.55 | 4.55 | 4.98 | 4.89 |
| | | 5 | 4 | 4.48 | 4.28 | 4.34 | 4.21 | | | 5 | 4 | 4.00 | 4.00 | 4.49 | 4.38 |
| | | 6 | 4 | 4.00 | 4.00 | 4.09 | 4.06 | | | 6 | 4 | 4.24 | 4.22 | 4.56 | 4.48 |
| | | 7 | 3 | 3.00 | 3.00 | 3.83 | 3.51 | | | 7 | 4 | 5.10 | 5.03 | 5.15 | 5.09 |
| | | 8 | 4 | 4.65 | 4.55 | 4.90 | 4.72 | | | 8 | 4 | 4.00 | 4.00 | 4.92 | 4.85 |
| | | 9 | 4 | 4.08 | 4.04 | 4.43 | 4.29 | | | 9 | 4 | 4.43 | 4.40 | 4.68 | 4.62 |
| | | 10 | 4 | 4.00 | 4.00 | 4.19 | 4.11 | | | 10 | 4 | 4.92 | 4.85 | 5.09 | 4.99 |
| | 0.2 | 1 | 8 | 8.44 | 8.21 | 8.91 | 8.39 | | 0.2 | 1 | 8 | 8.55 | 8.37 | 8.86 | 8.51 |
| | | 2 | 8 | 8.65 | 8.33 | 8.88 | 8.42 | | | 2 | 9 | 10.33 | 9.91 | 10.62 | 10.04 |
| | | 3 | 8 | 8.39 | 8.31 | 8.62 | 8.27 | | | 3 | 9 | 9.75 | 9.61 | 10.33 | 10.08 |
| | | 4 | 6 | 6.37 | 6.28 | 7.19 | 6.66 | | | 4 | 8 | 8.46 | 8.38 | 9.09 | 8.69 |
| | | 5 | 7 | 7.31 | 7.23 | 7.84 | 7.30 | | | 5 | 8 | 8.67 | 8.57 | 8.82 | 8.53 |
| | | 6 | 8 | 9.47 | 9.33 | 9.75 | 9.38 | | | 6 | 8 | 9.55 | 9.12 | 9.55 | 9.18 |
| | | 7 | 7 | 7.00 | 7.00 | 7.78 | 7.42 | | | 7 | 9 | 10.01 | 9.91 | 10.50 | 10.23 |
| | | 8 | 7 | 7.66 | 7.63 | 7.99 | 7.57 | | | 8 | 9 | 9.11 | 9.09 | 9.64 | 9.29 |
| | | 9 | 8 | 8.17 | 8.09 | 8.68 | 8.25 | | | 9 | 8 | 8.83 | 8.77 | 9.08 | 8.86 |
| | | 10 | 7 | 7.26 | 7.20 | 7.63 | 7.31 | | | 10 | 10 | 10.89 | 10.78 | 11.22 | 10.83 |

Para analizar la existencia de diferencias significativas aplicamos contrastes de hipótesis no paramétricos siguiendo el procedimiento propuesto en [5] y [12]. En primer lugar, para cada una de las instancias se calculan los rangos de los cinco algoritmos considerados A*, A, A1,

A2 y A12 ordenados de más a menos efectivo según el valor promedio de las 100 ejecuciones de cada uno de ellos. Se asigna 1 al mejor, 2 al segundo y así sucesivamente. Si hay empates se asigna el valor medio a todos los algoritmos empatados. Se calculan los valores medios de todos los rangos correspondientes a cada algoritmo. Estos rangos promedios son:

| A* | A | A1 | A2 | A12 |
|---|---|---|---|---|
| 1.53 | 2.78 | 2.17 | 4.89 | 3.63 |

Para las comparaciones múltiples entre los cinco algoritmos, tomando el algoritmo A* como referencia (tratamiento de control), se aplica en primer lugar el test de Friedman. Dado que el valor práctico del estadístico (261.01) es mucho mayor que el valor crítico de la distribución $F$ con 4 y 480 grados de libertad (2.39 al 5% y 3.32 al 1%) se deduce que existen diferencias significativas entre los rendimientos de los algoritmos.

Descartada la equivalencia de los algoritmos procedemos a realizar los contrastes a posteriores de Nemenyi para detectar entre qué par de algoritmo se encuentran las diferencias significativas. En primer lugar se ordenan de mejor a peor los rangos promedios de los cuatro algoritmos:

| A* | A1 | A | A12 | A2 |
|---|---|---|---|---|
| 1.53 | 2.17 | 2.78 | 3.63. | 4.89 |

Las diferencias observadas son:

| A1 − A* | A − A1 | A12 −A | A2 −A12 |
|---|---|---|---|
| 0.64 | 0.61 | 0.83 | 1.26 |

El valor del error típico es SE = $\sqrt{\frac{k(k+1)}{6n}} = 0.144$ de donde las diferencias críticas para el test de Nemenyi son $(3.089/\sqrt{2}) \cdot 0.144 = 0.56$ y $(3.978/\sqrt{2}) \cdot 0.144 = 0.72$ ya que 3.089 y 3.978 son los correspondientes valores de los rangos *studentizados* a esos niveles para 5 variables. Por tanto, las diferencias del algoritmo A1 tanto con el algoritmo A como con el algoritmo A* son significativas al 5% pero no al 1%.

Para analizar la capacidad de exploración de los algoritmos se han determinado el número de veces que se obtiene la solución óptima en las 100 repeticiones de cada algoritmo. Estos valores se recogen en la tabla III. En ella se destacan las instancias en los que a los algoritmos A y A1 les cuesta encontrar la solución óptima, y se encuentra con mayor probabilidad con los algoritmos A2 y A12. Obsérvese que usando el test de los signos de Wilcoxon con 240 instancias, las veces en que un algoritmo tiene que ser superior a otro para que las diferencias sean significativas es de 136 al 5% y 140 al 1%. Por tanto a estos niveles las diferencias no llegan a ser aún significativas

En la tabla IV se compara el rendimiento de estos algoritmos en un contexto de arranque múltiple. Esta tabla contiene los valores alcanzados por el algoritmo A* y las tres versiones de la estrategia piloto junto al mejor resultado de las 100 ejecuciones de cada los cuatro versiones del algoritmo MVCA. Se observa que la versión del algoritmo A12, el algoritmo MVCA revisado y aleatorizado, con post-optimización y primer color aleatorio consigue alcanzar el óptimo tras 100 repeticiones en todos los casos excepto en 1.

En cuanto a los tiempos de ejecución, excepto para los problemas mayores, se trata de unos pocos milisegundos por lo que muchos de los valores alcanzados carecen de valor teniendo en cuenta las limitaciones de los medidores de tiempos de los PC. La tabla III solo muestra los tiempos en milisegundos para $d = 0.2$ que es donde los algoritmos tardan más.

TABLA II
TIEMPO DE EJECUCIÓN

| $n$ | 20 | 30 | 40 | 50 |
|---|---|---|---|---|
| **A*** | 11.0 | 138.0 | 1002.0 | 66325.0 |
| A1* | 4.7 | 6.2 | 20.4 | 28.2 |
| A2* | 14.1 | 15.6 | 28.1 | 29.6 |
| A12* | 40.6 | 56.3 | 131.4 | 318.7 |
| A | 4.8 | 17.1 | 33.1 | 43.7 |
| A1 | 14.0 | 25.0 | 32.9 | 46.8 |
| A2 | 10.9 | 18.7 | 29.7 | 43.6 |
| A12 | 14.0 | 17.3 | 31.3 | 57.7 |

TABLA III
COMPARACIÓN CON LA ESTRATEGIA PILOTO

|  | $n = l = 20$ | | | $n = l = 30$ | | | $n = l = 40$ | | | $n = l = 50$ | | |
|---|---|---|---|---|---|---|---|---|---|---|---|---|
| $D$ | 0.8 | 0.5 | 0.2 | 0.8 | 0.5 | 0.2 | 0.8 | 0.5 | 0.2 | 0.8 | 0.5 | 0.2 |
| **A*** | 2.4 | 3.1 | 6.7 | 2.8 | 3.7 | 7.4 | 2.9 | 3.7 | 7.4 | 3.0 | 4.0 | 8.6 |
| A1* | 2.4 | 3.2 | 6.7 | 2.8 | 3.7 | 7.4 | 2.9 | 3.7 | 7.6 | 3.0 | 4.1 | 8.6 |
| A2* | 2.5 | 3.5 | 6.7 | 2.8 | 3.7 | 7.4 | 2.9 | 3.9 | 7.7 | 3.0 | 4.2 | 8.6 |
| A12* | 2.4 | 3.1 | 6.7 | 2.8 | 3.7 | 7.4 | 2.9 | 3.7 | 7.5 | 3.0 | 4.1 | 8.6 |
| A | 2.5 | 3.5 | 6.7 | 2.8 | 3.7 | 7.4 | 2.9 | 3.9 | 7.7 | 3.0 | 4.2 | 8.6 |
| A1 | 2.4 | 3.3 | 6.7 | 2.8 | 3.7 | 7.4 | 2.9 | 3.8 | 7.5 | 3.0 | 4.1 | 8.6 |
| A2 | 2.4 | 3.1 | 6.7 | 2.8 | 3.7 | 7.4 | 2.9 | 3.7 | 7.5 | 3.0 | 4.1 | 8.6 |
| A12 | 2.4 | 3.1 | 6.7 | 2.8 | 3.7 | 7.4 | 2.9 | 3.7 | 7.4 | 3.0 | 4.1 | 8.6 |

TABLA IV
NUMERO DE OPTIMOS ALCANZADOS POR CADA ALGORITMO

| N | d | i | A1 | A2 | A3 | A4 |
|---|---|---|----|----|----|----|
| 20 | 0.8 | 1 | 100 | 100 | 92 | 97 |
| | | 2 | **0** | **2** | **12** | **13** |
| | | 3 | 100 | 100 | 25 | 51 |
| | | 4 | 25 | 27 | 15 | 24 |
| | | 5 | 100 | 100 | 15 | 34 |
| | | 6 | 100 | 100 | 38 | 63 |
| | | 7 | 100 | 100 | 100 | 100 |
| | | 8 | 100 | 100 | 15 | 23 |
| | | 9 | 100 | 100 | 93 | 96 |
| | | 10 | 100 | 100 | 94 | 96 |
| | 0.5 | 1 | **0** | **5** | **10** | **13** |
| | | 2 | 100 | 100 | 40 | 78 |
| | | 3 | 100 | 100 | 31 | 63 |
| | | 4 | 56 | 66 | 29 | 38 |
| | | 5 | 48 | 57 | 25 | 33 |
| | | 6 | **0** | **20** | **21** | **28** |
| | | 7 | 100 | 100 | 100 | 100 |
| | | 8 | **0** | **0** | **7** | **10** |
| | | 9 | 100 | 100 | 22 | 45 |
| | | 10 | **0** | **0** | **9** | **9** |
| | 0.2 | 1 | 100 | 100 | 33 | 69 |
| | | 2 | 36 | 43 | 16 | 39 |
| | | 3 | 36 | 40 | 15 | 34 |
| | | 4 | 46 | 100 | 34 | 99 |
| | | 5 | 74 | 76 | 19 | 50 |
| | | 6 | 80 | 89 | 48 | 83 |
| | | 7 | 93 | 94 | 52 | 83 |
| | | 8 | 99 | 99 | 60 | 88 |
| | | 9 | 79 | 91 | 55 | 91 |
| | | 10 | 64 | 69 | 23 | 51 |
| 30 | 0.8 | 1 | 100 | 100 | 89 | 96 |
| | | 2 | 100 | 100 | 9 | 34 |
| | | 3 | 100 | 100 | 96 | 99 |
| | | 4 | 100 | 100 | 100 | 100 |
| | | 5 | 100 | 100 | 92 | 96 |
| | | 6 | 100 | 100 | 95 | 100 |
| | | 7 | 100 | 100 | 17 | 26 |
| | | 8 | 100 | 100 | 85 | 93 |
| | | 9 | 100 | 100 | 100 | 100 |
| | | 10 | 100 | 100 | 96 | 100 |
| | 0.5 | 1 | 100 | 100 | 85 | 93 |
| | | 2 | 86 | 87 | 70 | 77 |
| | | 3 | 59 | 61 | 12 | 34 |
| | | 4 | 100 | 100 | 29 | 50 |
| | | 5 | 52 | 72 | 66 | 79 |
| | | 6 | 100 | 100 | 91 | 94 |
| | | 7 | 100 | 100 | 18 | 49 |
| | | 8 | 35 | 45 | 18 | 29 |
| | | 9 | 92 | 96 | 58 | 71 |
| | | 10 | 100 | 100 | 81 | 89 |
| | 0.2 | 1 | 58 | 79 | 26 | 63 |
| | | 2 | 35 | 67 | 30 | 60 |
| | | 3 | 61 | 69 | 41 | 74 |
| | | 4 | 75 | 80 | 15 | 50 |
| | | 5 | 69 | 77 | 31 | 70 |
| | | 6 | **4** | **5** | **2** | **5** |
| | | 7 | 100 | 100 | 28 | 61 |
| | | 8 | 34 | 37 | 17 | 47 |
| | | 9 | 83 | 91 | 41 | 78 |
| | | 10 | 74 | 80 | 41 | 70 |

| n | d | i | A1 | A2 | A3 | A4 |
|---|---|---|----|----|----|----|
| 40 | 0.8 | 1 | 100 | 100 | 79 | 85 |
| | | 2 | 100 | 100 | 74 | 78 |
| | | 3 | 100 | 100 | 71 | 76 |
| | | 4 | 100 | 100 | 95 | 96 |
| | | 5 | 100 | 100 | 76 | 84 |
| | | 6 | 100 | 100 | 8 | 39 |
| | | 7 | 100 | 100 | 91 | 95 |
| | | 8 | 100 | 100 | 85 | 97 |
| | | 9 | 100 | 100 | 60 | 69 |
| | | 10 | 100 | 100 | 77 | 89 |
| | 0.5 | 1 | **0** | **1** | **4** | **5** |
| | | 2 | 64 | 68 | 60 | 63 |
| | | 3 | 74 | 75 | 44 | 51 |
| | | 4 | 100 | 100 | 73 | 90 |
| | | 5 | 100 | 100 | 32 | 53 |
| | | 6 | 100 | 100 | 29 | 43 |
| | | 7 | 100 | 100 | 5 | 25 |
| | | 8 | **0** | **0** | **7** | **14** |
| | | 9 | 100 | 100 | 55 | 64 |
| | | 10 | 65 | 65 | 12 | 27 |
| | 0.2 | 1 | 60 | 81 | 14 | 32 |
| | | 2 | 61 | 68 | 9 | 48 |
| | | 3 | **0** | **0** | **2** | **3** |
| | | 4 | **0** | **7** | **4** | **11** |
| | | 5 | **0** | **9** | **0** | **2** |
| | | 6 | 13 | 21 | 7 | 16 |
| | | 7 | 23 | 25 | 14 | 22 |
| | | 8 | 66 | 72 | 22 | 44 |
| | | 9 | 76 | 79 | 31 | 57 |
| | | 10 | 30 | 31 | 9 | 23 |
| 50 | 0.8 | 1 | 100 | 100 | 75 | 80 |
| | | 2 | 100 | 100 | 51 | 54 |
| | | 3 | 100 | 100 | 61 | 69 |
| | | 4 | 100 | 100 | 81 | 84 |
| | | 5 | 100 | 100 | 78 | 81 |
| | | 6 | 100 | 100 | 74 | 78 |
| | | 7 | 100 | 100 | 67 | 77 |
| | | 8 | 100 | 100 | 67 | 70 |
| | | 9 | 100 | 100 | 72 | 75 |
| | | 10 | 100 | 100 | 76 | 81 |
| | 0.5 | 1 | 100 | 100 | 68 | 73 |
| | | 2 | 100 | 100 | 28 | 47 |
| | | 3 | **0** | **0** | **0** | **0** |
| | | 4 | 45 | 45 | 9 | 15 |
| | | 5 | 100 | 100 | 51 | 62 |
| | | 6 | 76 | 78 | 44 | 52 |
| | | 7 | **0** | **3** | **2** | **2** |
| | | 8 | 100 | 100 | 10 | 17 |
| | | 9 | 57 | 60 | 32 | 38 |
| | | 10 | **8** | **15** | **3** | **7** |
| | 0.2 | 1 | 50 | 64 | 30 | 54 |
| | | 2 | **4** | **19** | **7** | **24** |
| | | 3 | 44 | 54 | 17 | 27 |
| | | 4 | 55 | 63 | 22 | 50 |
| | | 5 | 33 | 43 | 31 | 48 |
| | | 6 | **2** | **15** | **3** | **11** |
| | | 7 | 16 | 22 | 6 | 13 |
| | | 8 | 89 | 91 | 45 | 72 |
| | | 9 | 31 | 34 | 17 | 27 |
| | | 10 | 22 | 29 | 13 | 24 |

## IV. CONCLUSIONES

Hemos propuesto dos modificaciones tendentes a mejorar la capacidad de explotación y exploración del algoritmo MVCA revisado para el MSTP: un proceso de post-optimización y un paso inicial aleatorio, respectivamente. Se ejecutaron 100 veces cada una de las versiones sobre un conjunto de 120 instancias obtenidas de la literatura especializada. Los contrastes estadísticos no paramétricos de Friedman y de Nemenyi permiten sustentar la conclusión de que la post-optimización consigue aumentar significativamente la capacidad de explotación. Los contrastes de los signos de Wilcoxon para sustentar la conclusión de que el arranque aleatorio del MVCA revisado (con o sin post-optimización) consigue mejorar la capacidad de exploración del mismo. Sin embargo la incorporación de ambas modificaciones a la vez no mejora la capacidad de explotación del algoritmo.

Como investigación futura se analizará el rendimiento de un procedimiento GRASP de arranque múltiple que consiga aprovechar la mejora de la explotación del post-optimización y la de la exploración del arranque aleatorio. Para ello, el método GRASP propuesto arrancará en las primeras iteraciones con la estrategia del MVCA revisado, para en las siguientes repeticiones utilizar un arranque aleatorio. Este procedimiento se comparará también con una VNS y otras propuestas de la literatura en instancias de mayor tamaño y en extensiones del MLSTP inspiradas en situaciones reales, como son la consideración de un problema de tipo Steiner o de problemas donde las aristas pueden tener varios colores o etiquetas de forma simultánea

## REFERENCIAS

[1] Brüggemann T., Monnot J., Woeginger G. J., (2003). 'Local search for the minimum label spanning tree problem with bounded color classes', *Operations Research Letters*, 31: 195–201.

[2] Cerulli R., Fink A., Gentili M., Voss S., (2005). 'Metaheuristics comparison for the minimum labelling spanning tree problem'. In Golden B.L, Raghavan S. and Wasil E.A. (eds.), *The Next Wave on Computing, Optimization, and Decision Technologies*, New York: in Springer, 93 - 106.

[3] Chang R.S., Leu S.J., (1997). 'The minimum labeling spanning trees', *Information Processing Letters* 63(5): 277–282.

[4] Consoli, S., Moreno, J.A., Mladenovic, N., Darby-Dowman, K. (2006). 'Constructive Heuristics for the Minimum Labelling Spanning Tree Problem: a preliminary comparison'. DEIOC Documentos de Trabajo 2006/4, Available at: http://webpages.ull.es/users/estinv/Investigacion/pdfs_dt/DT_DEIOC_4_2006.pdf, and at http://bura.brunel.ac.uk/handle/2438/504.

[5] Demšar J., (2006). 'Statistical Comparison of Classifiers over multiple Data Sets', *Journal of Machine Learning Research* 7:1-30.

[6] Duin C., Voss S., (1999). 'The pilot method: A strategy for heuristic repetition with applications to the Steiner problem in graphs', *Wiley InterScience*, Vol. 34 (3): 181-191.

[7] Friedman M., (1937). 'The use of ranks to avoid the assumption of normality implicit in the analysis of variance', *Journal of the American Statistical Association*, 32:675–701.

[8] Friedman. M., (1940). 'A comparison of alternative tests of significance for the problem of m rankings', *Annals of Mathematical Statistics*, 11:86–92.

[9] Golden B., Raghavan S., Stanojević, (2004). 'Heuristic search for the generalized minimum spanning tree problem'. *INFORMS J. Computation XXX*

[10] Iman R.L., Davenport J.M., (1980). 'Approximations of the critical region of the Friedman statistic', *Communications in Statistics*, pp. 571–595.

[11] Krumke S.O., Wirth H.C., (1998). 'On the minimum label spanning tree problem', *Information Processing Letters* 66(2): 81–85.

[12] Moreno, J.A., Campos, C. Laguna, M. (2007) 'Introducción a las técnicas estadísticas no paramétricas para la comparación de metaheurísticas'. MAEB 2007, Tenerife.

[13] Sedgewick R., (2002). *'Algorithm in C – Part 5: Graph algorithms'*, Addison-Wesley.

[14] Shaffer J.P., (1995). 'Multiple hypothesis testing', *Annual Review of Psychology*, 46:561–584.

[15] Tanenbeum A.S., (1989). 'Computer Networks', Prentice-Hall.

[16] Voss S., Cerulli R., Fink A. and Gentili M., (2005). 'Applications of the pilot method to hard modifications of the minimum spanning tree problem', *18th MINI EURO Conference on VNS*, Tenerife, Spain.

[17] Wan Y., Chen G. and Xu Y., (2002). 'A note on the minimum label spanning tree', Information Processing Letters 84: 99–101.

[18] Xiong Y., Golden B. and Wasil E., (2002). 'Worst case behavior of the MVCA heuristic for the minimum labeling spanning tree problem', *Operations Research Letters* 33(1):77–80.

[19] Xiong Y., Golden B. and Wasil E., (2005). 'A One-Parameter Genetic Algorithm for the Minimum Labeling Spanning Tree Problem', *IEEE Transactions on Evolutionary Computation.*, vol. 9, no. 1.

[20] Xiong Y., (2005).*'The Minimum Labeling Spanning Tree Problem and some variants'*, Ph.D. Thesis, graduate School of the University of Maryland, USA.